\documentstyle[graphicx,fancyhdr,times,epsf]{acmconfJOB}
\begin{document}
\newcommand{\PSbox}[3]{\mbox{\rule{0in}{#3}\includegraphics{#1}\hspace{#2}}}
\raggedbottom

\setcounter{page}{71}

\pagestyle{fancy}
\renewcommand{\footrulewidth}{0pt}
\thispagestyle{fancy}
\fancyfoot{}
\fancyhead{}

\rfoot{\ifodd\thepage{\thepage}\else\relax\fi}
\lfoot{\ifodd\thepage\relax\else\thepage\fi}
\lhead{\ifodd\thepage\relax\else\fontsize{8}{8}\textsf{ACM SIGGRAPH 95, Los Angeles, California, August 6--11,1995}\fi}
\rhead{\ifodd\thepage\fontsize{8}{8}\textsf{Computer Graphics Proceedings, Annual Conference Series, 1995}\else\relax\fi}

\newlength{\headrulelength}
\setlength{\headrulelength}{\textwidth}
\newlength{\headrulegap}
\setlength{\headrulegap}{.75in}
\addtolength{\headrulelength}{-\headrulegap}
\def\headrule{\mbox{\hspace{\headrulegap}\rule[2ex]{\headrulelength}{\headrulewidth}}\gdef\headrule{\hrule}}

\fancypagestyle{empty}{
  \fancyfoot{}
  \fancyhead{}
  \fancyhead[RO,LE]{\fontsize{8}{8}\textsf{Computer Graphics Proceedings, Annual Conference Se
ries, 1995}}
  \fancyfoot[RO,LE]{\thepage}
  \fancyhead[LO,RE]{\includegraphics[height=.66in]{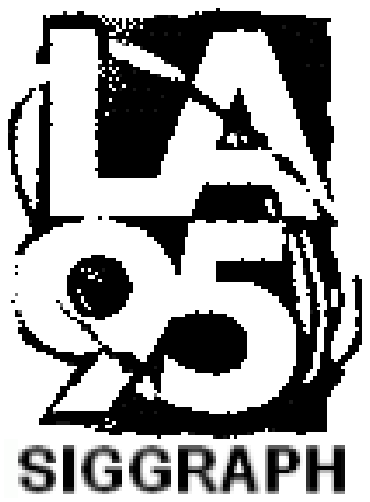}\vspace*{-.46in}}
}

\title{Animating Human Athletics}
\author{Jessica K. Hodgins 
\item Wayne L. Wooten 
\item David C. Brogan 
\item James F. O'Brien}
\affiliation{College of Computing, Georgia Institute of Technology}
\maketitle
\footnotetext{College of Computing, Georgia Institute of Technology,
Atlanta, GA 30332-0280. [jkh$|$wlw$|$dbrogan$|$obrienj]@cc.gatech.edu}
\pretolerance=900\tolerance=1200

\section*{ABSTRACT}

This paper describes algorithms for the animation of men and women 
performing three dynamic athletic behaviors: 
running, bicycling, and vaulting.  We animate these behaviors
using control algorithms 
that cause a physically realistic model to perform the desired maneuver. 
For example, control algorithms allow the simulated humans to maintain 
balance while moving their arms, to run or bicycle at a variety of 
speeds, and to perform a handspring vault.  Algorithms for 
group behaviors allow a number of simulated bicyclists to ride as a group 
while avoiding simple patterns of obstacles.  We add secondary motion 
to the animations with spring-mass simulations of clothing driven by the 
rigid-body motion of the simulated human.  For each simulation, we compare
the computed motion to that of humans performing similar 
maneuvers both qualitatively through the comparison of real and 
simulated video images and quantitatively through the comparison of 
simulated and biomechanical data. 

\vskip 1em
\noindent
{\bf Key Words and Phrases:} computer animation, human motion, motion control, 
dynamic simulation, physically realistic modeling.

\section*{INTRODUCTION}

People are skilled at perceiving the subtle details of human motion.
We can, for example, often identify friends by the style of their walk when they are 
still too far away to be recognizable otherwise.  If synthesized human 
motion is to be compelling, we must create actors for 
computer animations  and virtual environments that appear 
realistic when they move.  Realistic human motion has several components: 
the kinematics and dynamics of the figure must be physically correct and 
the control algorithms must make the figure perform in ways that 
appear natural.  We are interested in the last of these: the
design of control strategies for natural-looking human motion.  

\begin{figure}[tb]
\begin{center}
\PSbox{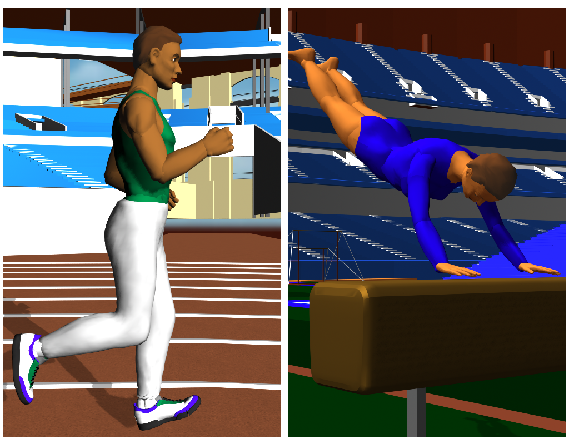} {2.5in}{1.8in}
\end{center}
\vskip -.15in
{\small
\begin{caption}
{Images of a runner on the track in the 1996 Olympic Stadium
and a gymnast performing a handspring vault in the Georgia Dome.}
\label{fig:photo}
\end{caption}
}
\end{figure}
\footnotetext{College of Computing, Georgia Institute of Technology,
Atlanta, GA 30332-0280. [jkh$|$wlw$|$dbrogan$|$obrienj]@cc.gatech.edu\\
    {\small\textsf{%
Permission to make digital/hard copy of part or all of this work for
personal or classroom use is granted without fee provided that copies
are not made or distributed for profit or commercial advantage, the
copyright notice, the title of the publication and its date appear, and
notice is given that copying is by permission of ACM, Inc. To copy
otherwise, to republish, to post on servers, or to redistribute to
lists, requires prior specific permission and/or a fee.
    \parbox[t][0in][b]{\columnwidth}{%
      \small\textsf{\vspace*{-3ex}\copyright 1995\hfill ACM-0-89791-701-4/95/008\hfill \$3.50}}\vspace*{-6ex}
 }}}

In particular, this paper describes algorithms that allow a rigid-body model 
of a man or woman to stand, run, and turn at a variety of speeds, 
to ride a bicycle on hills and around obstacles, and to perform a gymnastic 
vault.  
Figure~\ref{fig:photo} shows two examples of the animated
\pagebreak
behaviors.
The rigid-body models of the man and woman are realistic in 
that their mass and inertia properties 
are derived from data in the biomechanics literature and 
the degrees of freedom of the joints are chosen so that each 
behavior can be completed in a natural-looking fashion.

Although the behaviors are very different in character, the control 
algorithms are built from a common toolbox: 
state machines are used to enforce a correspondence between the 
phase of the behavior and the active control laws, 
synergies are used 
to cause several degrees of freedom to act with a single purpose,
limbs without required actions in a particular state are used to 
reduce disturbances to the system,
inverse kinematics is 
used to compute the joint angles that would cause a foot or hand to 
reach a desired location, and the low-level control is performed with 
proportional-derivative control laws.  

We have chosen to animate running, bicycling, and vaulting because 
each behavior contains a 
significant dynamic component.  For these behaviors, the dynamics of the 
model constrain the motion and limit the space that must 
be searched to find control laws for natural-looking motion.  This property 
is most evident in the gymnastic vault.  The gymnast is airborne for 
much of the maneuver, and the control algorithms can influence the internal 
motion of the joints but not the angular momentum of the system as a whole.
The runner, on the other hand, is in contact with the ground much of the time 
and the joint torques computed by the control algorithms directly control 
many of the details of the motion.  Because the dynamics do not
provide as many constraints on the motion, much more effort went
into tuning the motion of the runner to look natural than into tuning 
the motion of the gymnast.

\pagebreak
Computer animations and interactive virtual environments require 
a source of human motion.  The approach used here, dynamic simulation 
coupled with control algorithms, is only one of several options. An 
alternative choice, motion capture, is now widely available in commercial 
software.  The difficulty of designing control algorithms has 
prevented the value of simulation from being demonstrated for systems with 
internal sources of energy, such as humans.  However,
simulation has several potential 
advantages over motion capture.  Given robust control algorithms, simulated 
motion can easily be computed to produce similar but different motions 
while maintaining physical realism (running at $\rm 4~m/s$ rather than 
$\rm 6~m/s$ 
for example).  Real-time simulations also allow the motion of an animated 
character to be truly interactive, an important property for virtual 
environments in which the actor must move realistically in response to 
changes in the environment and in response to the actions of the user.  And 
finally, when the source of motion is dynamic simulation we have the 
opportunity 
to use multiple levels of simulation to generate either secondary motion 
such as the movement of clothing and hair or high-level motion such as 
obstacle avoidance and group behaviors.


\section*{BACKGROUND}

Research in three fields is relevant to the problem of animating
human motion:  robotics, biomechanics, and computer graphics.
Researchers in robotics have explored control techniques for legged
robots that walk, run, balance, and perform gymnastic maneuvers.  
While no robot has been built with a complexity similar to that of 
the human body, control strategies for simpler machines provide 
basic principles that can be used to design control strategies for 
humanlike models.

Raibert and his colleagues built and controlled a series of dynamic 
running machines, ranging from a planar machine with one telescoping
leg to three-dimensional machines that ran on two or
four legs.  These machines walked, jumped, changed gait, climbed stairs, 
and performed gymnastic maneuvers ([14--16], [26--28]).
The control algorithms for human 
running described in this paper build on these control algorithms 
by extending them for systems with many more controlled degrees of freedom 
and more stringent requirements on the style of the motion.

Biomechanics provides the data and hypotheses about human motion required 
to ensure that the computed motion resembles that of a human performing 
similar maneuvers.  The biomechanics literature contains motion capture data, 
force plate data, and muscle activation records for many human behaviors.  
These data were used to tune the control algorithms for running, bicycling, and
balancing.  Cavagna presents energy curves for walking and running as well 
as studies of energy usage during locomotion\cite{Cavagna}.
McMahon provides graphs of stance duration, flight duration, and step 
length as a function of forward speed\cite{McMahon}. Gregor
surveys biomechanical studies of bicyclists\cite{Gregor}. 
Takei presents biomechanical data of elite female 
gymnasts performing a handspring vault and relates the data to
the scores that the gymnasts received in competition\cite{Takei}.

Many researchers in computer graphics have explored the difficult problems
inherent in animating human motion.  
The Jack system developed at the University of Pennsylvania contains 
kinematic and dynamic models of humans based on biomechanical 
data\cite{Badler}.
It allows the interactive positioning of the body 
and has several built-in behaviors including 
balance, reaching and grasping, and walking and running 
behaviors that use 
generalizations of motion capture data\cite{Ko}.

Bruderlin and Calvert used a simplified dynamic model and control 
algorithms to generate the motions of a walking human\cite{Bruderlin}.
The leg model included
a telescoping leg with two degrees of freedom for the stance phase and 
a compound pendulum model for the swing phase.  A foot, upper body, and 
arms were added to the model kinematically, and were made to move in 
an oscillatory pattern similar to that observed in humans.  
Pai programmed a walking behavior for a dynamic model of a 
human torso and legs in a high-level fashion by describing a
set of time-varying constraints, such as, ``maintain ground clearance 
during leg swing,'' ``lift and put down a foot,'' ``keep the torso vertical,'' 
and ``support the torso with the stance leg''
\cite{Pai}.

None of these approaches to generating motion for animation are automatic
because each new behavior requires additional work on the part of the 
researcher.  In recent years, the field has seen the development of a 
number of techniques for automatically generating motion for new behaviors 
and new creatures.  Witkin and Kass\cite{Witkin},
Cohen\cite{Cohen}, and Brotman and Netravali\cite{Brotman} treat the 
problem of automatically generating motion as a trajectory optimization 
problem.  Another approach finds a control 
algorithm instead of a desired trajectory (\cite{vandePanne90},
\cite{vandePanne93}, \cite{Ngo}, \cite{Sims94a}, and \cite{Sims94b}).
In contrast, the control algorithms described in this paper 
were designed by hand, using a toolbox of control techniques, our physical 
intuition about the behaviors, observations of humans performing the tasks, 
and biomechanical data.  While automatic techniques would be preferable to 
hand design, automatic techniques have not yet been developed that can find 
solutions for systems with the number of controlled degrees of freedom needed 
for a plausible model of the human body.  Furthermore, although the motion 
generated by automatic techniques is appealing, much of it does not appear 
natural in the sense of resembling the motion of a biological system. We do 
not yet know whether this discrepancy is because only relatively simple 
models have been used or because of the constraints and optimization 
criteria that were chosen.

\section*{DYNAMIC BEHAVIORS}

The motion of each behavior described in this paper is computed using
dynamic simulation.  Each simulation contains the equations of motion 
for a rigid-body model of a human and environment (ground, bicycle,
and vault), control algorithms for balancing, running, 
bicycling, or vaulting, a graphical image for viewing the motion, and a user 
interface for changing the parameters of the simulation.  The user is provided 
with limited high-level control of the animation.  For example, the desired 
velocity and facing direction for the bicyclist and runner are selected by the 
user.  During each simulation time step, the control algorithm computes desired
positions and velocities for each joint based on the state of the system, 
the requirements of the task and input from the user.  
Proportional-derivative servos compute joint torques based on the 
desired and actual value of each joint.  The equations of motion of the system 
are integrated forward in time taking into account the internal joint 
torques and the external forces and torques from interactions with the 
ground plane or other objects.  The details of the human model and the 
control algorithm for each behavior are described below.

\begin{figure}[tb]
\centerline{\epsfxsize=2.0in \epsfbox{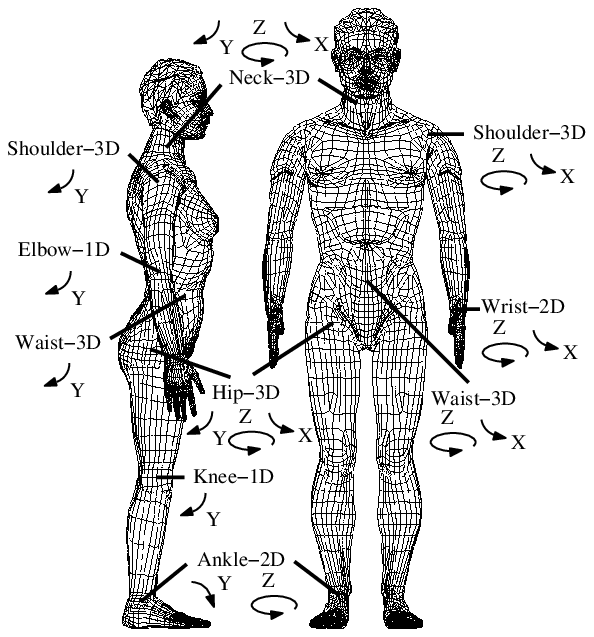}}
\vskip -.1in
\begin{caption}
{The controlled degrees of freedom of the human model.  The gymnast 
represented in the figure has
15 body segments and a total of 30 controlled degrees of freedom.
The runner has 17 body segments and 30 controlled 
degrees of freedom (two-part feet with a one degree of freedom
joint at the ball of the foot and only one degree of freedom at the ankle),
The bicyclist has 15 body segments and 22 controlled degrees of freedom 
(only one degree of freedom 
at the neck, hips, and ankles).  The directions of the arrows indicates
the positive direction of rotation for each degree of freedom. The polygonal 
models were purchased from Viewpoint Datalabs. 
}\label{fig:manwomanmodel}
\end{caption}
\end{figure}

\subsection*{Human Models}

The human models we used to animate the dynamic behaviors were constructed from 
rigid links connected by rotary joints with one, two or three degrees of 
freedom.  The dynamic models were derived from the graphical models 
shown in figure~\ref{fig:manwomanmodel} by computing the mass 
and moment of inertia of each body part using 
algorithms for computing the moment of inertia 
of a polygonal object of uniform density\cite{Lien} and
density data measured from cadavers\cite{Dempster}.
We also verified that the model could 
perform maneuvers that rely on the parameters of the dynamic system using 
data from Frohlich\cite{Frohlich}.

The controlled degrees of freedom of the models are shown in 
figure~\ref{fig:manwomanmodel}. Each internal joint of 
the model has a very simple muscle model, a torque source, that allows 
the control algorithms to apply a torque between the two links that form 
the joint.  The equations of motion for each system 
were generated using a commercially available package\cite{Rosenthal}.
The points of contact between the feet and the ground, the 
bicycle wheels and the ground, and the gymnast's hands and the vault are 
modeled using constraints.  The errors for the constraints are the 
relative accelerations, velocities, and positions of one body with respect 
to the other.  The constraints are stabilized using Baumgarte 
stabilization\cite{Baumgarte}.  

\begin{figure}[t]
\begin{center}
\vskip -.25in
\PSbox{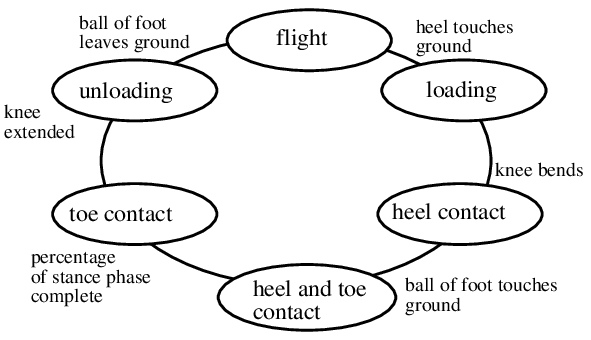} {2.5in}{1.5in}
\end{center}
\vskip -.25in
\begin{caption}
{A state machine is used to determine the control actions that 
should be active for running given the current state of the system.  
The transition 
events are computed for the active leg. At liftoff the active and idle 
legs switch roles.  The states correspond to the points of contact on the 
ground: flight, heel contact, heel and
toe contact, and toe contact.  The other two states, loading and unloading,
are of very short duration and ensure that the foot is firmly planted on the
ground or free of the ground before the control actions for stance or flight
are invoked.}
\label{fig:statemachine}
\end{caption}
\end{figure}

\subsection*{Running}

Running is a cyclic behavior in which the legs swing fore and aft and provide
support for the body in alternation.  Because the legs perform different 
functions during the phases of the locomotion cycle, the muscles are used 
for different control actions at different times in the cycle.  When the 
foot of the simulated runner is on the ground, the ankle, knee, 
and hip provide support and balance.  During the flight phase, a leg 
is swung forward in preparation for the next touchdown.  These distinct phases 
and corresponding changes in control actions make a state machine a natural 
tool for selecting the control actions that should be active at a 
particular time.  The state machine and transition events used for the 
simulation of running are shown in figure~\ref{fig:statemachine}.

To interact with the animation of the runner, the user specifies 
desired values for the magnitude of the velocity on the ground plane 
and the facing direction.  The control laws for each state 
compute joint torques that move the velocity and facing direction 
toward these desired values while maintaining balance.  The animated runner
can run at speeds between 2.5~m/s and 5~m/s and runs along a user-defined path.

We call the leg that is on the ground or actively being positioned for 
touchdown the {\it active leg.}  The other leg is called the {\it idle leg}. 
During flight, the active leg is swung forward in anticipation of touchdown.  
Using the degrees of freedom of the leg in a synergistic fashion,
the foot is positioned at touchdown to correct for errors in forward
speed and to maintain balance.  Forward speed is controlled by placing 
the average point of support during 
stance underneath the hip and taking into account the change in
contact point from heel to metatarsus during stance.  
At touchdown, the desired distance from the hip to the heel 
projected onto the ground plane is 
\begin{eqnarray}
x_{\rm hh} & = & 1/2 (t_{\rm s} \dot x - \cos (\theta) l_{\rm f}) + k
(\dot x - \dot x_{\rm d}) \\
y_{\rm hh} & = & 1/2 (t_{\rm s} \dot y - \sin (\theta) l_{\rm f}) + k
(\dot y - \dot y_{\rm d})
\end{eqnarray}
where $t_{\rm s}$ is an estimate of the period of time that the foot will 
be in contact with the ground (based on the previous stance duration), 
$\dot x$ and $\dot y$ are the velocities of the runner on the plane,
$\dot x_{\rm d}$ and $\dot y_{\rm d}$ are the desired velocities, 
$\theta$ is
the facing direction of the runner, $l_{\rm f}$ is the distance
from the heel to the ball of the foot, and $k$ is a gain for the 
correction of errors in speed. The length of the leg at touchdown is 
fixed and is used to calculate the vertical distance from the 
hip to the heel, $z_{\rm hh}$.  The disturbances caused by the impact 
of touchdown can be reduced by decreasing the relative speed 
between the foot and the ground at touchdown.  This technique is called
{\it ground speed matching} in the biomechanical literature.
In this control system, ground speed matching is accomplished
by swinging the hip further forward in the direction of travel during flight
and moving it back just before touchdown.  

The equations for $x_{\rm hh}$, $y_{\rm hh}$, and $z_{\rm hh}$, and the
inverse kinematics of the leg are used to compute the desired knee 
and hip angles at touchdown for the active leg.  The angle of the ankle 
is chosen so that the toe will not touch the ground at the same time 
as the heel at the beginning of stance.

During stance, the knee acts as a passive spring to store the kinetic 
energy that the system had at touchdown. The majority of the vertical 
thrust is provided by the ankle joint.  During the first part of stance,
{\it heel contact}, the toe moves toward the ground because the contact 
point on the heel is behind the ankle joint.  
Contact of the ball of the foot triggers the 
transition from {\it heel contact} to {\it heel and toe contact.}  The transition
from {\it heel and toe contact} to {\it toe contact} occurs when the time
since touchdown is equal to a percentage of the expected stance
duration (30-50\% depending on forward speed).  After the transition, 
the ankle joint is extended, causing the heel to lift off the ground and 
adding energy to the system for the next flight phase.

Throughout stance, proportional-derivative servos are used to compute torques 
for the hip joint of the stance leg that will cause the attitude of the body 
(roll, pitch, and yaw) to move toward the desired values. The desired angle 
for roll is zero except during turning when the body leans into the curve.  
The desired angle for pitch is inclined slightly forward, and the desired 
angle for yaw is set by the user or the higher-level control algorithms
for grouping behaviors and obstacle avoidance.

\pagebreak
The idle leg plays an important role in locomotion by reducing
disturbances to the body attitude caused by the active leg as it
swings forward and in toward the centerline in preparation for touchdown.  
The idle leg is shortened so that the toe does not stub the ground, and 
the hip angles mirror the motion of the active leg to reduce the net
torque on the body:
\begin{eqnarray}
\alpha_{\rm x_d} = \alpha_{\rm x_{lo}} - (\beta_{\rm x_{d}}
	- \beta_{\rm x_{lo}}) \\
\alpha_{\rm y_d} = \alpha_{\rm y_{lo}} - (\beta_{\rm y_{d}}
	- \beta_{\rm y_{lo}}) 
\end{eqnarray}
where $\alpha_{\rm x_d}$ and $\alpha_{\rm y_d}$ are the 
desired rotations of the idle hip with respect to the pelvis,
$\alpha_{\rm x_{lo}}$ and $\alpha_{\rm y_{lo}}$ 
are the rotation of the idle hip at the previous liftoff, $\beta_{\rm x_d}$ 
and $\beta_{\rm y_d}$ are the desired position of the active hip, 
and $\beta_{\rm x_{lo}}$ and $\beta_{\rm y_{lo}}$ are the 
position of the active hip at the previous liftoff. 
The mirroring action of the idle leg is modified by the restriction 
that the legs should not collide as they pass each other during stance.

The shoulder joint swings the arms fore and aft in a motion that is 
synchronized with the motion of the legs:
\begin{equation}
\gamma_{\rm y_d }= k \alpha_{\rm y} + \gamma_0
\end{equation}
where $\gamma_{\rm y_d}$ is the desired fore/aft angle for the shoulder,
$k$ is a scaling factor,
$\alpha_{\rm y}$ is the fore-aft hip angle for the leg on the opposite 
side of the body, and $\gamma_0$ is an offset.  The other two degrees of 
freedom in the shoulder ($x$ and $z$) and the elbows also follow a 
cyclic pattern with the same period as $\gamma_{\rm y_d}$.
The motion of the upper body is important in running because the counter
oscillation of the arms reduces the yaw oscillation of the body
caused by the swinging of the legs.  However, the details of
the motion of the upper body are not constrained by the dynamics of 
the task and amateur athletes use many different styles of arm motion
when they run.  Observations of human runners were used to tune the 
oscillations of the arms to produce a natural-looking gait.

The control laws compute desired values for each joint and 
proportional-derivative servos are used to control the position
of all joints.  For each internal joint the control equation is
\begin{equation}
\tau = k (\theta_d - \theta) +  k_v (\dot \theta_d - \dot \theta)
\end{equation}
The desired values used in the proportional-derivative servos 
are computed as trajectories from the current value of 
the joint to the desired value computed by the control laws.  
Eliminating large step changes in the errors 
used in the proportional-derivative servos smoothes the simulated motion.

\begin{figure}[tb]
\centerline{\epsfxsize=2.0in \epsfbox{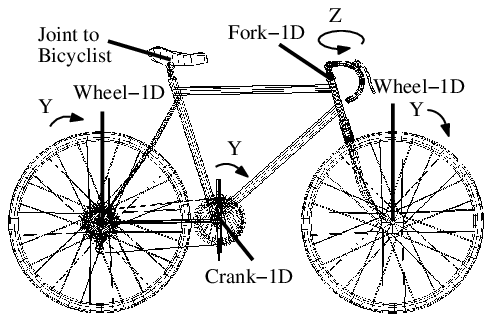}}
\vskip -.1in
\begin{caption}
{The four degrees of freedom of the bicycle model. The direction
of the arrows indicates the positive direction of rotation for each
degree of freedom. The polygonal model
is a modification of a model purchased from Viewpoint Datalabs.
}\label{fig:bicycleangles}
\end{caption}
\end{figure}

\subsection*{Bicycling}

The bicyclist controls the facing direction and speed of the 
bicycle by applying forces to the handlebars and pedals
with his hands and feet.  The rider is attached to the bicycle by a pivot 
joint between the bicycle seat and the pelvis 
(figure~\ref{fig:bicycleangles}).  
Spring and damper systems connect the hands to the handlebars, the feet 
to the pedals, and the crank to the rear wheel.  The connecting springs 
are two-sided and the bicyclist is able to pull up on the pedals as if 
the bicycle were equipped with toe-clips and a fixed gear (no freewheel).  
The connection 
between the crank and the rear wheel includes an adjustable gear ratio.  
The bicycle wheels have a rolling resistance proportional to the velocity.

The control algorithm adjusts the velocity of the bicycle by using the legs
to produce a torque at the crank.  The desired torque at the crank is
\begin{equation}
\tau_{\rm crank} = k (v - v_d)
\end{equation}
where $k$ is a gain, $v$ is the magnitude of the bicyclist's velocity, and $v_d$ 
is the desired velocity.  The force applied by each leg depends on the 
angle of the crank because we assume that the legs are most 
effective at pushing downwards.  For example, the front leg can 
generate a positive torque and the rear leg can generate a negative torque
when the crank is horizontal.
To compensate for the crank position, the desired forces for the legs are 
scaled by a weighting function between zero and one that depends on the 
crank position, $\theta_{\rm crank}$:
\begin{equation}
w = {\sin (\theta_{\rm crank}) + 1  \over 2}.
\end{equation}
$\theta_{\rm crank}$ is zero when the crank is vertical and the
right foot is higher than the left.  If $\tau_{\rm crank} > 0$, the 
force on the pedal that the legs should produce is
\begin{eqnarray}
f_l & = & {w \tau_{\rm crank} \over l} \\
f_r & = & {(1 - w) \tau_{\rm crank} \over l}
\end{eqnarray}
where $f_l$ and $f_r$ are the desired forces from the left and right
legs respectively, and $l$ is the length of a crank arm. 
If $\tau_{\rm crank}$ is less than zero, then the equations for the left 
and right leg are switched.  An inverse kinematic model of the legs is used to 
compute hip and knee torques that will produce the desired pedal forces.

To steer the bicycle and control the facing direction, the control algorithm
computes a desired angle for the fork based on the errors in roll and yaw:
\begin{equation}
\theta_{\rm fork} = -k_{\alpha} (\alpha - \alpha_d) - k_{\dot \alpha} 
	\dot \alpha + k_{\beta} (\beta - \beta_d) + k_{\dot \beta} \dot \beta
\end{equation}
where $\alpha$, $\alpha_d$, and $\dot \alpha$ are the roll angle, desired 
roll, and roll velocity and
$\beta$, $\beta_d$, and $\dot \beta$ are the yaw angle, desired yaw,
and yaw velocity.  $k_{\alpha}$, $k_{\dot \alpha}$, $k_\beta$, 
and $k_{\dot \beta}$ are gains.  The desired yaw angle is set by the user
or high-level control algorithms;
the desired roll angle is zero.  Inverse kinematics is used to compute the 
shoulder and elbow angles that will position the hands on the handlebars 
with a fork angle of $\theta_{\rm fork}$.  Proportional-derivative 
servos move the shoulder and elbow joints toward those angles.

These control laws leave the motion of several of the joints of the bicyclist 
unspecified.  The wrists and the waist are held at a nearly constant angle with 
proportional-derivative controllers.  The ankle joints are controlled to 
match data recorded from human subjects\cite{CavanaghSanderson}.

\section*{\bf Vaulting and Balancing}

To perform a vault, the gymnast uses a spring board to launch herself 
toward the vaulting horse, pushes off the horse with her hands, and lands
on her feet on the other side of the horse.  The vault described
here, a handspring vault, is one in which 
the gymnast performs a full somersault over the horse while keeping her 
body extended in a layout position. This vault is structured by a state
machine with six states: hurdle step, board contact, first flight, horse 
contact, second flight, and landing.  The animation of the handspring 
vault begins 
during the flight phase preceding the touchdown on the springboard.
The initial conditions were estimated from video footage
(forward velocity is 6.75~m/s and the height of the center of mass is 0.9~m). 

The simulated gymnast lands on a spring board that 
deflects based on a linear spring and damper model.  When the 
springboard reaches maximum deflection, the control system extends the
knees, pushing on the springboard and adds energy to the system. As 
the springboard rebounds, it launches the gymnast into the air and the 
first flight state begins.  Using a technique called {\it blocking},
the control system positions the hips forward 
before touchdown on the springboard so that much of the horizontal 
velocity at touchdown is transformed into rotational and vertical 
velocity at liftoff.

During the first flight state, the control system prepares to put the
gymnast's hands on the horse by positioning her shoulders on the line
between the shoulders and the desired hand position on the vault:
\begin{equation}
\gamma_{\rm y_d} = \lambda_{\rm y} - \phi
\end{equation}
where $\gamma_{\rm y_d}$ is the desired shoulder angle relative to the body,
$\lambda_{\rm y}$ is the angle between vertical and a vector from the 
shoulder to the desired hand position on the vault, and $\phi$ is the 
pitch angle of the body (with respect to vertical).  Because the
shoulders are moving toward the vault during flight, this control law 
performs ground speed matching between the hands 
and the horse.  The wrists are controlled to cause the hands to hit the 
horse palm down and parallel to the surface of the horse. 

During the next state, the gymnast's hands contact the vault and the arms 
are held straight.  No torque is applied at the shoulder or the wrist and the
angular and forward velocity of the gymnast carries her over the horse as she 
performs the handspring. When the hands leave the vault, the second flight phase 
begins.

During the second flight state, the control system maintains a layout position 
with the feet spread slightly to give a larger area of support at touchdown.  
When the feet hit the ground, the control system
must remove the horizontal and rotational energy from the somersault 
and establish an upright, balanced position.  
The knees and waist are bent to absorb 
energy.  Vaulters land on soft, 4~cm thick mats that help to reduce their 
kinetic energy.  In our simulation, the behavior of the mat is approximated
by reducing the stiffness of the ground.
When the simulated gymnast's center of mass passes over the center 
of the polygon formed by the feet, a balance controller is activated.
After the gymnast is balanced, the control system straightens the knees 
and hips to cause the gymnast to stand up. 

The balance controller not only allows the gymnast to stand up after
a landing but also compensates for disturbances resulting from the motion of 
other parts of the body while she is standing.  For example, if the gymnast 
bends forward, the ankles are servoed to move the center of mass of
the gymnast backwards.  The balance controller also
allows the simulated gymnast throw her arms back in a gesture of success
after the vault and to take a bow.

\section*{\bf HIGHER-LEVEL BEHAVIORS}

\begin{figure*}[p]
\begin{center}
\PSbox{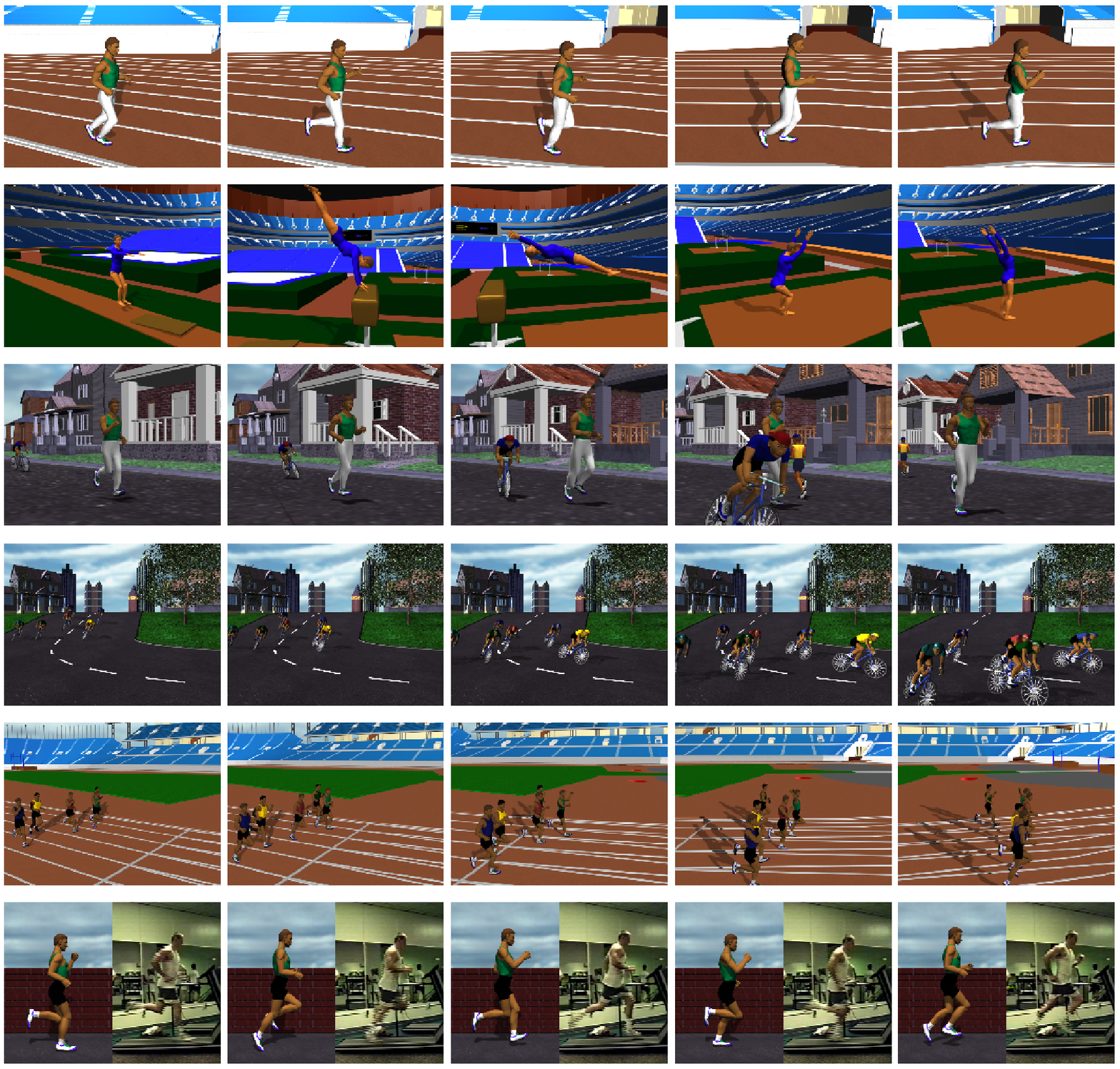} {6.75in}{6.5in}
\end{center}
{\small
\begin{caption}
{Images of an athlete wearing sweat pants running on a quarter mile track 
in the 1996 Olympic Stadium, 
a gymnast performing a handspring vault in the Georgia Dome,
a bicyclist avoiding a jogger, 
a group of bicyclists riding around a corner during a race,
a group of runners crossing the finish line,
and a comparison between a simulated and a human runner on a treadmill. 
In each case, the spacing of the images in time is equal with 
the stadium runner at intervals of 0.066~s, 
the gymnast at 0.5~s, 
the single bicyclist at 1.0~s,
the group of bicyclists at 0.33~s,
the group of runners at 0.5~s
and the composite of the simulated and human runner at 0.066~s.
}
\label{fig:filmstrip}
\end{caption}
}
\end{figure*}

The control algorithms provide the animator with control over the 
velocity and facing direction of the runner and bicyclist.  However, 
choreographing an animation with many bicyclists or runners would 
be difficult because the animator must ensure that they do not 
run into each other while moving as a group and avoiding obstacles.
Building on Reynolds\cite{Reynolds}, we implemented an algorithm that 
allows bicyclists to move as a group and to avoid simple configurations 
of obstacles on the terrain.  The performance of the algorithm for a
simulation of a bike race on a hill is shown in figure~\ref{fig:filmstrip}.

In contrast to most previous implementations of algorithms for group
behaviors, we use this algorithm to control a group where the
members have significant dynamics.  The problem of controlling these
individuals more closely resembles that faced by 
biological systems because each individual has limited 
acceleration, velocity, and turning radius.  Furthermore, the control 
algorithms for bicycling are inexact, resulting both in transient and 
steady-state errors in the control of velocity and facing direction.  

The algorithm for group behaviors computes a desired position for each
individual by averaging the location and velocity of its visible neighbors, 
a desired group velocity, and a desired position with respect to
the visible obstacles.  
The details of this computation are presented in Brogan and 
Hodgins\cite{BroganHodgins}.
This desired position is then used as an input to the control algorithm 
for the bicyclist.  The desired position is known only to the individual 
bicyclist and his navigational intent is communicated to the 
other cyclists only through their observation of his actions. 

The desired position for the bicycle that is computed by the algorithm for
group behaviors is used to compute a desired velocity and facing direction:
\begin{equation}
v_{\rm d} = k_{\rm p} e + k_{\rm v} (v_{\rm gl} -  v)
\end{equation}
where $v_{\rm d}$ is the desired velocity in the plane,
$v$ is the actual velocity,
$e$ is the error between the current position of the bicyclist and 
the desired position, $k_{\rm p}$ is the proportional gain on position,
$k_{\rm v}$ is the proportional gain on velocity, and $v_{\rm gl}$ is 
the group's global desired velocity (specified by the user).   

\section*{\bf SECONDARY MOTIONS}

While we are often not consciously aware of secondary motions, they can
add greatly to the perceived realism of an animated scene. This
property is well known to traditional animators, and much of the work
in creating believable hand animation focuses on animating the motion
of objects other than the primary actors. This effect can be duplicated
in computer animation by identifying the objects in the environment
that should exhibit passive secondary behavior and including a
simulation suitable for modeling that type of behavior.  In some cases,
the simulated motion of the passive secondary objects can be driven by
the rigid body motion of the primary actors.  As examples of this
approach, we have simulated sweatpants and splashing water.  The
behavior of the sweatpants is computed by using the motion of the
simulated runner to drive a passive system that approximates the
behavior of cloth.  Similarly, the motion of splashing water is driven
by the motion of a platform diver when it impacts the
water (\cite{OBrien} and \cite{Wooten}).  
Ideally, all objects that do not have active
control could be implemented in this fashion. Unfortunately,
computational resources and an incomplete understanding of physical
processes restrict the size and types of the passive systems that we
are able to simulate.

Several methods for physically based animation of cloth have been
described in the literature (\cite{Carignan}, \cite{MagnenatThalmann},
\cite{Breen}, and \cite{Terzopoulos}).  Carignan\cite{Carignan}
implemented a system that uses the motion of a kinematic human walker
developed by Laurent\cite{Laurent} to drive the action of the cloth.
Our approach is similar to that described by Terzopoulos and
Fleischer\cite{Terzopoulos}.  We use an elastic model to define the
properties of the cloth.  Collisions are detected using a hierarchical
object grouping algorithm  and resolved using inverse dynamics to
compute reaction forces.  Although our cloth model is not significantly
different from previous methods,  our approach of using dynamically
correct rigid body motion to drive the passive system results in an
animated scene where all the motion is governed by a consistent set of
physically based rules.


\section*{\bf DISCUSSION}

This paper presents algorithms that allow an animator to generate motion for 
several dynamic behaviors.  
Animations of platform diving, unicycle 
riding and pumping a swing have been described elsewhere 
(\cite{Wooten}, \cite{HodginsSweeney}).  
Taken together with previous work, these dynamic behaviors represent a 
growing library.  While these behaviors do not
represent all of human motion or even of human athletic endeavors, an 
animation package with ten times this many behaviors would have sufficient 
functionality to be interesting to students and perhaps even to professional 
animators.

Several open questions remain before the value of simulation as a source of 
motion for animation and virtual environments can be conclusively demonstrated:  

How can we make it easier to generate control algorithms for a new behavior? 
This paper partially addresses that question by presenting a toolbox of 
techniques that can be used to construct the control algorithms for a 
set of diverse behaviors.  However, developing sufficient physical 
intuition about a new behavior to construct a robust control algorithm 
remains time consuming.  We hope that these examples represent a 
growing understanding of the strategies that are useful in controlling 
simulations of human motion and that this understanding 
will lead to the development of more automatic techniques.

What can we do to reduce the number of new behaviors that need to be 
developed?  One idea that has been explored by researchers in the domain of
motion capture and keyframing is to perform transitions between behaviors 
in an automatic or semiautomatic fashion.  Such transitions may be much more
amenable to automatic design than the design of entire control algorithms
for dynamic simulations.

What rules can we add to the system to improve the naturalness of the motion?  
The techniques presented here are most effective for behaviors with
a significant dynamic component because the dynamics constrain the number 
of ways in which the task can be accomplished.
When the gross characteristics of the motion are not constrained by the 
dynamics of the system, the task can be completed successfully but 
in a way that appears unnatural.  For example, the simulated runner can run 
while holding his arms fixed at his sides, but an animation of that motion 
would be amusing rather than realistic.  Humans are strong enough and
dextrous enough that simple arm movements such as picking up a coffee cup
can be completed in many different ways. In contrast, only good
athletes can perform a handspring vault and the variations seen in their 
performances are relatively small.  When the dynamics do not significantly 
constrain the task, the control algorithms must be carefully designed and 
tuned to produce motion that appears natural while matching the key features 
of the behavior when performed by a human.  
The tuning process might be aided by data from psychophysical experiments
that would provide additional constraints for the motion.

Can human motion be simulated interactively? To be truly interactive, the 
motion of synthetic actors in virtual environments must be computed in 
real time (simulation time must be less than wall clock time). Our 
implementation of the bicyclist runs ten times slower than real time on a 
Silicon Graphics $\rm{Indigo}^2$ Computer with an R4400 processor.  We 
anticipate that with improved dynamic simulation techniques, 
and the continued increase in workstation speed, a three-dimensional human 
simulation will run in real time within a few years.

A related question is whether the behaviors are robust enough for the 
synthetic actors to interact in a natural fashion with unpredictable 
human users.  
The runner can run at a variety of speeds and change direction, but abrupt 
changes in velocity or facing direction will cause him to fall down.  The 
planning or reactive response algorithms that lie between the locomotion 
control algorithms and the perceptual model of the simulated environment 
will have to take in account the limitations of the dynamic system and 
control system.



\begin{figure}[tb]
\hskip 0in \PSbox{./ps/simhipknee.ps} {1.5in}{1.65in}
\vskip -1.65in \hskip 1.70in \PSbox{./ps/humhipknee.ps}{1.5in}{1.65in}
\vskip -.1in
{\small
\begin{caption}
{A phase plot of the hip and knee angles seen in the simulated runner (left) and
measured in human subjects (right).  The simulated motion is qualitatively 
similar to the measured data.}
\label{fig:hipkneedata}
\end{caption}
}
\end{figure}

One goal of this research is to demonstrate that dynamic simulation of
rigid-body models can be
used to generate natural-looking motion.  Figure~\ref{fig:filmstrip} shows 
a side-by-side comparison of video footage of a human runner and images
of the simulated runner.  This comparison represents one form of evaluation 
of our success in generating natural-looking motion.
Figure~\ref{fig:hipkneedata} shows biomechanical
data for running and represents another form of validation. 
Table~\ref{table:comparison} compares data 
from female gymnasts\cite{Takei} and data from the vault simulation.


\begin{table}[t]
\begin{center}
\scriptsize
\begin{tabular} {|l|r|r|r|r|} \hline
Variables & \multicolumn{3}{c|}{Human} & \multicolumn{1}{c|}{Simulation} \\
& \multicolumn{1}{c}{Mean}  & \multicolumn{1}{c}{Min} &
\multicolumn{1}{c|}{Max} & \multicolumn{1}{c|}{} \\[.02in] \hline
Mass (kg) & 47.96 & 35.5 & 64.0 & 64.3 \\
Height (m) & 1.55 & 1.39 & 1.66 & 1.64 \\
Board contact (s) & 0.137 & 0.11 & 0.15 & 0.105 \\ 
First flight (s) & 0.235 & 0.14 & 0.30 & 0.156 \\
Horse contact (s) & 0.245 & 0.19 & 0.30 & 0.265 \\
Second flight (s) & 0.639 & 0.50 & 0.78 & 0.632 \\
Horizontal velocity (m/s) &  & & &  \\
\quad Board touchdown & 6.75 & 5.92 & 7.25 & 6.75 \\
\quad Board liftoff & 4.61 & 3.97 & 5.26 & 4.01 \\
\quad Horse touchdown & 4.61 & 3.97 & 5.26 & 4.01 \\
\quad Horse liftoff & 3.11 & 2.48 & 3.83 & 2.83 \\
Vertical velocity (m/s) &  & & &  \\
\quad Board touchdown & -1.15 & -1.54 & -.71 & -1.13 \\
\quad Board liftoff & 3.34 & 2.98 & 3.87 & 3.81 \\
\quad Horse touchdown & 1.26 & 0.74 & 2.39 & 2.13 \\
\quad Horse liftoff & 1.46 & 0.56 & 2.47 & 1.10 \\
Aver. vertical force (N) & & & &  \\
\quad Board contact & 2175 & 1396 & 2792 & 5075 \\
\quad Horse contact & 521 & 309 & 752 & 957 \\ \hline
\end{tabular} 
\end{center}
{\small
\vskip -.1in
\begin{caption}
{
Comparison of velocities, contact times, and forces for a simulated
vaulter and human data measured by Takei. The human data was 
averaged from 24 subjects. The simulated data was taken from a
single trial.
}
\label{table:comparison}
\end{caption}
}
\end{table}

From the perspective of computer graphics, the final test would be a 
form of the Turing Test.  If simulated data and motion capture data were 
represented using the same graphical model, would the audience occasionally
choose the simulated data as the more natural motion? The user may find 
it easy to identify the motion source because motion capture data often 
has noise and registration problems with limbs that appear to change 
length and feet that slide on the ground.  Simulated motion also has 
characteristic flaws, for example, the cyclic motion of the runner is 
repetitive allowing the eye to catch oscillations in the motion that 
are not visible in the motion of the human runner.  

The animations 
described in this paper and a Turing test comparison with motion 
capture data can be seen on the WWW at
\noindent
{\tt \bf http://www.cc.gatech.edu/gvu/animation/Animation.html}

\section*{\bf ACKNOWLEDGMENTS}

The authors would like to thank Debbie Carlson and Ron Metoyer
for their help in developing our simulation and rendering environment,
Jeremy Heiner and Tom Meyer for their modeling expertise,
Amy Opalak for digitizing the motion of the bicyclist,
John Snyder and 
the User Interface and Graphics Research group at Microsoft
for allowing the use of their collision detection system,
and the CAD Systems Department at the Atlanta Committee for the Olympic 
Games for allowing us
to use models of the Olympic venues. This project was supported 
in part by NSF NYI Grant No. IRI-9457621, by Mitsubishi Electric Research 
Laboratory, and by a Packard Fellowship.  Wayne Wooten was supported by a 
Intel Foundation Graduate Fellowship.

 \end{document}